\documentclass[conference]{IEEEtran}
\usepackage{graphicx}

\begin{document}

\pagenumbering{gobble}

\title{Critical slowing-down as indicator of approach to the loss of stability}

\author{\IEEEauthorblockN{Dmitry Podolsky}
\IEEEauthorblockA{Department of Mechanical Engineering,\\Massachusetts Institute of Technology\\
Cambridge, MA, 02139\\
Email: podolsky@mit.edu}
\and
\IEEEauthorblockN{Konstantin Turitsyn}
 \IEEEauthorblockA{Department of Mechanical Engineering,\\Massachusetts Institute of Technology\\
Cambridge, MA, 02139\\
Email: turitsyn@mit.edu}
}

\maketitle
\begin{abstract}
We consider stochastic electro-mechanical dynamics of an overdamped
power system in the vicinity of the saddle-node bifurcation associated
with the loss of global stability such as voltage collapse or phase
angle instability. Fluctuations of the system state vector are driven
by random variations of loads and intermittent renewable generation.
In the vicinity of collapse the power system experiences so-called
phenomenon of critical slowing-down characterized by slowing and simultaneous
amplification of the system state vector fluctuations. In generic
case of a co-dimension 1 bifurcation corresponding to the threshold
of instability it is possible to extract a single mode of the system
state vector responsible for this phenomenon. We characterize stochastic
fluctuations of the system state vector using the formal perturbative
expansion over the lowest (real) eigenvalue of the system power flow
Jacobian and verify the resulting expressions for correlation functions
of the state vector by direct numerical simulations. We conclude that
the onset of critical slowing-down is a good marker of approach to
the threshold of global instability. It can be straightforwardly detected
from the analysis of single-node autostructure and autocorrelation
functions of system state variables and thus does not require full
observability of the grid. 
\end{abstract}
%\begin{IEEEkeywords}
%Blackout prevention, emergency control, phasor measurements, power
%system stability, voltage stability, wide-area measurements and control. 
%\end{IEEEkeywords}

\section{Introduction}

Most of the US national power grid has ultimately
been shaped and built during the quick urbanization period of the
beginning of 20$^{{\rm th}}$ century. Increasing power demand continues
to put enormous strain on the infrastructure of the aging grid, forcing
utilities to maximally utilize its existing resources. As a consequence,
many parts of the grid often operate in nearly critical regimes with
significantly increased probability of large scale failures \cite{Andersson2005,Canizares2002,Pourbeik2006}.
In the modern world it becomes crucially important to be able to read
the state of the pre-critical smart grid in a timely and precise manner,
to effectively control parameters of smart grids operating in pre-critical
regimes minimizing the probability of their large scale failures.
Unfortunately, commonly used used power system state estimation and
control algorithms, working extremely well for stable operating regimes,
become less efficient (see for example \cite{Wu1977,Klump2000}) when
the operating point of the grid is close to the threshold of instability.

As we shall discuss below, a partial reason for the decrease of efficiency
of state estimation procedures is related to a strong amplification
of fluctuations of the power system state vector closer to the threshold
of instability. The present contribution addresses the problem of
stochastic fluctuations of power system state vector close to voltage
or phase-angle instabilities.%
\footnote{ The stochastic models of power system dynamics have been studied
before in a number of papers, such as \cite{Loparo1985,Nwankpa1992,Anghel2007,Dong2012,Odun-Ayo2012} %
}

The amplification of stochastic fluctuations of the system state vector
in pre-critical operating regimes is also directly related to the
phenomenon of critical slowing-down often observed in power grids
on the brink of failure \cite{Hines2011,Cotilla-Sanchez2012,Podolsky2012,Ghanavati2013}.
Naturally, detecting the onset of critical slowing-down by itself
can provide an efficient way for early detection of approach to a
large-scale instability \cite{Hines2011,Cotilla-Sanchez2012,Podolsky2012,Ghanavati2013}.
As will be explained below, the onset of critical slowing-down is
naturally associated with a strong growth of fluctuations of the system
state vector. This growth can in turn be locally identified well befiore
the event of collapse from local synchrophasor measurements of voltage
phase and magnitude on a given node of the smart grid. 

The contributions of our work can be summarized as follows. We provide
a formal mathematical description of critical slowing-down phenomenon,
and characterize it in terms of power flow Jacobian as well as generator
inertia and damping matrices. Second, we derive a closed form equation
for the autocorrelation and autostructure functions as well as power
spectral density of state vector that completely characterizes the
probabilities of arbitrary system trajectories. Remarkably, the autocorrelation
and autostructure functions are expressed in terms of steady-state
power flow Jacobian and its eigenvectors. Finally, we validate all
our results with numerical simulation of IEEE 39 and IEEE 57 test
systems.

%%%%%%%%%%%%%%%%%%%%%%%%%%%%%%%%%%%%%%%%%%%%%%%%%%%%%%%%%%%%%%%%
\section{Power flow and load models\label{sec:flow-load-model}}
%%%%%%%%%%%%%%%%%%%%%%%%%%%%%%%%%%%%%%%%%%%%%%%%%%%%%%%%%%%%%%%%%

To describe dynamics of system variables close to the threshold of
instability, we use the structure-preserving model \cite{Bergen1981}
which reduces to the system of coupled swing equations on $(P,V)$
(generator) nodes of the power grid 
%%%%%%%%%%%%%%%%%%%%%%%
\begin{equation}
\frac{H_{i}}{\pi f_{0}}\frac{d^{2}\theta_{i}}{dt^{2}}+\alpha_{i}\frac{d\theta_{i}}{dt}=\sum_{j\sim i}\mathcal{Y}_{ij}V_{i}V_{j}\sin(\theta_{i}-\theta_{j}-\gamma_{ij})+P_{m,i}
\label{eq:swing}
\end{equation}
%%%%%%%%%%%%%%%%%%%%%%%%
and power flow equations on $(P,Q)$ (consumer) nodes of the grid
%%%%%%%%%%%%%%%%%%%%%%%%%%%
\begin{equation}
P_{i}\approx P_{0,i}+\alpha_{p,i}\dot{\theta}_{i}+\beta_{p,i}(V_{i}-V_{0})+T_{p,i}\dot{V}_{i}=
\label{eq:P}
\end{equation}
 %%%%%%%%%%%%%%%%%%%%%%%%%%%
\[
=\sum_{j\sim i}{\cal Y}_{ij}V_{i}V_{j}\sin(\theta_{i}-\theta_{j}-\gamma_{ij}),
\]
 %%%%%%%%%%%%%%%%%%%%%%%%%%%
\begin{equation}
Q_{i}\approx Q_{0,i}+\alpha_{q,i}\dot{\theta}_{i}+\beta_{q,i}(V_{i}-V_{0})+T_{q,i}\dot{V}_{i}=
\label{eq:Q}
\end{equation}
 %%%%%%%%%%%%%%%%%%%%%%%%
\[
=\sum_{j\sim i}{\cal Y}_{ij}V_{i}V_{j}\cos(\theta_{i}-\theta_{j}-\gamma_{ij}).
\]
 %%%%%%%%%%%%%%%%%%%%%%%%
Here $\theta_{i}$ is a voltage phase on a bus $i$, $H_{i}$ is inertia
constant of a generator on the node $i$, parameters $\alpha_{i}$
describe droop controls of generators and/or load dependence on frequency
fluctuations $\omega_{i}=\dot{\theta}_{i}$ on the $(P,Q)$ nodes,
$V_{i}$ is a voltage magnitude on a bus $i$ (for the $(P,V)$ nodes,
$V_{i}=E_{i}$), parameters $\beta$ encode dependence of the power
load on the voltage magnitude $V_{i}$. Finally, coefficients $T_{i}$
describe a (weak) dependence of the power load on the rate $\dot{V}_{i}$
of the voltage change with time. The power losses are important for
the dynamics of system state variables, thus generally $\gamma_{ij}\ne0$.

\begin{figure}[tbh]
\includegraphics[scale=0.5]{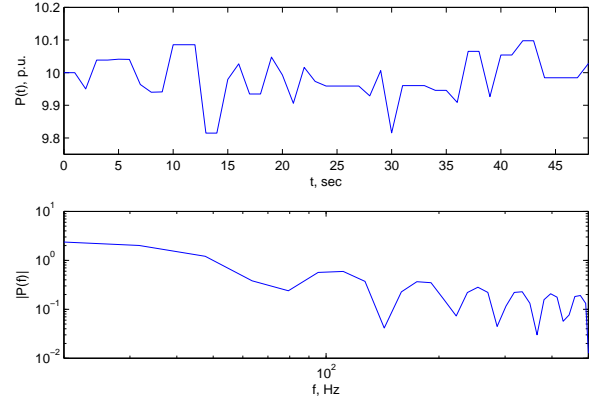} 
\protect\protect\protect\caption{Time profile of aggregated power load $P_{i}(t)$ and the absolute
value of its Fourier transform. 
\label{fig:load-time-profile}}
\end{figure}

We consider the load model (\ref{eq:P}), (\ref{eq:Q}) with such
parameters chosen that $\beta_{i}=0$ for simplicity, the choice of
$\alpha_{i}$ corresponds to $1\div2$\% change of the load per $1$\%
change in system frequency $f$ and $10\div20$\% change in power
generation per $1$\% change in $f$ \cite{Kundur1994}, while parameters
$T_{i}$ --- to $0.1\%$ change in power load per $1\%$ change in
$\dot{V}$. For simplicity we assume, that all the loads have fixed
power factors $PF_{i}=k_{i}/\sqrt{1+k_{i}^{2}}$, and all fluctuate
with time in the vicinity of the average (base) load $\bar{P_{i}}(t)$,
constantly deviating from the base value and returning back, so that
$P_{i}(t)=\bar{P}_{i}(t)+\delta P_{i}(t),$ $Q_{i}(t)=\bar{Q}_{i}(t)+\delta Q_{i}(t)$
\cite{Loparo1985,Nwankpa1992}. The same applies to generated power,
especially if the grid is exposed to intermittence of renewable energy
sources. Thus, $P_{i}$ generally behave as stochastic processes \cite{Cox1977}.
The base load value $\bar{P}_{i}(t)$ itself changes with time but
relatively slowly, with significant changes only noticeable at time
scales of several hours. In principle, a given, aggregated load $P_{i}(t)$
can be thought of as an aggregation of a huge number of power-consuming
devices connected to the node $i$ of the grid, which get online and
offline, connected and disconnected from the grid. A resulting overall
time profile of the aggregated active load $P_{i}(t)$ is represented
on the Fig. \ref{fig:load-time-profile}. The characteristic time
scale $t_{{\rm on/off}}$ of step-like changes depicted there is $0.1-1$
sec. 
%%%%%%%%%%%%%%%%%%%%%%%%%%%%%%%%%%%%%%%%%
%%%%%%%%%%%%%%%%%%%%%%%%%%%%%%%%%%%%%%%%%%%%
These relatively fast power load fluctuations on different nodes of
the grid are statistically independent from each other, and their
correlation properties of the vectors $\delta P$, $\delta Q$ can
be considered Gaussian \cite{Cox1977}. The latter have mean zero
and are thus completely characterized by the expectation values 
%%%%%%%%%%%%%%%%%%%%%%%%%%
\begin{equation}
E(\delta P(t)\delta P^{T}(t'))={\cal B}(t-t'),
\label{eq:Pcorr}
\end{equation}
 %%%%%%%%%%%%%%%%%%%%%%%%%%
while 
%%%%%%%%%%%%%%%%%%%%%%%%%%%%%
\[
E(\delta Q(t)\delta Q^{T}(t'))=k{\cal B}(t-t')k^{T},
\]
 %%%%%%%%%%%%%%%%%%%%%%%%%%%
and 
%%%%%%%%%%%%%%%%%%%%%%%%%%%
\[
E(\delta P(t)\delta Q^{T}(t'))=k{\cal B}(t-t'),
\]
%%%%%%%%%%%%%%%%%%%%%%%%%%
here ${\cal B}(t-t')$ is a diagonal covariance matrix. Note that
generally ${\cal B}(t-t')\ne{}{\rm Const.}$ Alternatively, the fluctuations
can be represented with the power spectral density $E(\delta P(f)\delta P^{T}(-f))$,
which is simply a Fourier transform of the autocorrelation function
(\ref{eq:Pcorr}), see for example \cite{Cox1977}. Here and in the
forthcoming paragraphs the averaging corresponds to summation over
possible realization of the load fluctuations. For stationary stochastic
processes, the same results could be recovered by averaging over long
time periods $T$. 
 
Whenever the random load profile is composed of a continuing sequence
of independent individual ``jumps'', the power spectral density
${\cal B}(f)$ behaves like ${\cal B}_{0}/((2\pi f)^{2}+\tau^{-2})$,
where $\tau$ is the characteristic time of power load correlations.
We assume that this time is much longer than the typical timescales
of $1-30s$ that are considered in this work, but smaller than the
scale of several hours. On the time scales correspoding to large enough
frequencies, such that $f\gg\tau^{-1}$, one approximately has ${\cal B}(f)={\cal B}_{0}/(2\pi f)^{2}$.
The same approximation has been also proposed and justified in \cite{Hauer2007}.
Importantly, this model is different from the ones considered in a
number of other studies that rely on a more traditional ``white noise''
model \cite{Odun-Ayo2012,Ghanavati2013,Wang2013}. The latter implies
that the correlation properties of the fluctuating power loads are
described by $E(\delta P(f)\delta P(-f))\approx{\cal B}_{0}$, independent
of $f$. The power of such fluctuations of $\delta P$ does not diminish
with frequency, which is not realistic.

Close to the equilibrium operating point, the nonlinear system of
equations (\ref{eq:swing}), (\ref{eq:P}), (\ref{eq:Q}) can be linearized:
%%%%%%%%%%%%%%%%%%%%%%%%%%%%%%%%%%%%%%%%%%%%%%%%
\begin{equation}
{\cal M}\ddot{x}+{\cal D}\dot{x}+{\cal K}x=\delta P,
\label{eq:swinglinearized}
\end{equation}
 %%%%%%%%%%%%%%%%%%%%%%%%%%%%%%%%%%%%%%%%%%%%%%%
where $x$ is the system state vector respreseting the deviations
of voltage phases and magnitudes from the equilibrium value, ${\cal K}$
is the power flow Jacobian, the diagonal matrix ${\cal M}$ encodes
inertia of the turbines, while the diagonal matrix ${\cal D}$ corresponds
to the frequency damping. The system (\ref{eq:swinglinearized}) of
Stochastic Differential Equations (SDE in what follows) will be the
main subject of our study.

%%%%%%%%%%%%%%%%%%%%%%%%%%%%%%%%%%%%%%%%%%%%%%%%%%%%%%%%%%%%%%%%%%%%%%%%%%%%%%%%%%%
\section{Stochastic dynamics of system variables near the threshold of instability:
theoretical description \label{sec:General-case-autocorrelation}}
%%%%%%%%%%%%%%%%%%%%%%%%%%%%%%%%%%%%%%%%%%%%%%%%%%%%%%%%%%%%%%%%%%%%%%%%%%%%%%%%%%%%

We discuss the general case of a power grid with a large number of
$(P,V)$ nodes (generators) and $(P,Q)$ nodes (loads). It is natural
to assume that the graph of the grid under consideration does not
have any specific translational/rotational symmetries, in which case
the saddle-node bifurcation, corresponding to the global collapse
of the power system, has a co-dimension 1. The autocorrelation function
$C(\delta{}t)=E((x(t+\delta{}t)x^{T}(t))$ of the system state vector
$x$ is given by 
%%%%%%%%%%%%%%%%%%%%%%%%%%%%%%
\begin{equation}
C(|\delta{}t|)=\int df\, e^{-j2\pi f|\delta{}t|}C(f),
\label{eq:CorrFourier}
\end{equation}
%%%%%%%%%%%%%%%%%%%%%%%%%%%%%%
where 
%%%%%%%%%%%%%%%%%%%%%%%%%%%%%%
\begin{equation}
C(f)=\frac{1}{(2\pi{}f)^{2}+\tau^{-2}}\cdot\mathcal{A}^{-1}(f)\mathcal{B}({\cal {\cal A}}^{\dagger}(f))^{-1}
\label{eq:AutoCorrFrequency}
\end{equation}
%%%%%%%%%%%%%%%%%%%%%%%%%%%%%%
and the system matrix is 
%%%%%%%%%%%%%%%%%%%%%%%%%%%%%%
\[
-\mathcal{A}(f)=\mathcal{M}(2\pi f)^{2}+j2\pi\mathcal{D}f+\mathcal{K}.
\]
%%%%%%%%%%%%%%%%%%%%%%%%%%%%%%
The value of the integral (\ref{eq:CorrFourier}) is determined by
the singularities of the integrand in the complex plane of $f$, which
in particular include zeros of the determinants $\det{\cal A}(f)$
and $\det{\cal A}(-f)$ as well as poles of $B(f)$ at $f=\pm{}(2\pi\tau)^{-1}$.
When the autocorrelation function of the system state vector in the
frequency domain is analyzed, the contribution of the the poles of
$B(f)$ can be filtered out by a high-pass filter. In time domain
one can study the autostructure function $S(|t-t'|)$ instead of $C(|t-t'|)$,
where the large contribution of the pole of $B(f)$ is cancelled out.

Naturally, among the remaining singularities, the one closest to the
real $f$ axis undermines behavior of the autostructure function (\ref{eq:CorrFourier})
at large $\delta{}t$. To identify this singularity, we recall that
near a co-dimension 1 saddle node bifurcation one (and only one) of
the eigenvalues of the power flow Jacobian ${\cal K}$ vanishes \cite{Dobson1992a}.
Close to the bifurcation, the inverse of the power flow Jacobian can
be written as 
%%%%%%%%%%%%%%%%%%%%%%%%%%%%%%
\begin{equation}
{\cal K}^{-1}=\frac{1}{\epsilon}ba^{T}+\tilde{{\cal K}}^{-1}\approx\frac{1}{\epsilon}ba^{T},
\label{eq:Krepresent}
\end{equation}
%%%%%%%%%%%%%%%%%%%%%%%%%%%%%%
where $\epsilon$ is the eigenvalue of ${\cal K}$ vanishing at the
bifurcation, while $a$ and $b$ are the corresponding left and right
eigenvectors. 

Under these assumptions, the leading singularity of the integrand
in (\ref{eq:CorrFourier}) coincides with a zero of $\det{\cal A}(f)$
(or $\det{\cal A}(-f)$ depending on the sign of the difference $t-t'$).
Such singularity is a simple pole by assumption that the center manifold
of the power system is one-dimensional. Constructing perturbation
theory in powers of small $\epsilon$ and assuming the overdamped
operating regime, one finds that the dominating mode determining behavior
of the expectation value (\ref{eq:CorrFourier}) at $|t-t'|\to\infty$
is given by 
%%%%%%%%%%%%%%%%%%%%%%%%%%%%%%%%%
\begin{equation}
2\pi f_{{\rm corr}}=-\frac{j\epsilon}{a^{T}{\cal D}b}
\label{eq:s0}
\end{equation}
%%%%%%%%%%%%%%%%%%%%%%%%%%%%%%%%%%
to the leading order in $\epsilon$. The ${\cal O}(\epsilon^{2})$
contribution can be neglected as long as $\epsilon\ll(a^{T}{\cal D}b)^{2}/(a^{T}{\cal D}^{T}\tilde{{\cal K}}^{-1}{\cal D}b-a^{T}{\cal M}b)$.
Note that the dependence of the leading mode on the matrix ${\cal M}$
of inertia constants appears only in the second order in $\epsilon$
and is negligible in the vicinity of the bifurcation point, as $\epsilon\to0$.
The frequency of the leading mode is purely imaginary to the order
${\cal {O}}(\epsilon)$ unless $\epsilon>(a^{T}{\cal D}b)^{2}/4a^{T}{\cal M}b$,
i.e., when the system is underdamped. 

Estimating the integral (\ref{eq:CorrFourier}) near the leading singularity
$f=f_{{\rm corr}}$, one finally finds to the leading order in $\epsilon$
%%%%%%%%%%%%%%%%%%%%%%%%%%%%%%%%%%%%%%%%%%
\begin{equation}
S(|\delta{}t|)=\frac{b(a^{T}{}Ba)b^{T}|\delta{}t|}{\epsilon^{2}}+
\label{eq:CorrFfinal}
\end{equation}
%%%%%%%%%%%%%%%%%%%%%%%%%%%%%%%%%%%%%%%%%%
\[
\frac{b(a^{T}{}Ba)(a^{T}{}Db)b^{T}}{\epsilon^{3}}\left(e^{-\frac{\epsilon{}|\delta{}t|}{a^{T}Db}}-1\right).
\]
%%%%%%%%%%%%%%%%%%%%%%%%%%%%%%%%%%%%%%%%%%%
Note that the matrix elements $a^{T}{\cal B}a$ and $a^{T}{\cal D}b$
are simple numerical factors, so that the dominating direction in
the phase space of the system where the fluctuations of the system
vector $x$ grow coincides with the direction of the right eigenvector
$b$ corresponding to the lowest eigenvalue $\epsilon$ of the power
flow Jacobian ${\cal K}$, \cite{Dobson1992a,Canizares1995}.

%%%%%%%%%%%%%%%%%%%%%%%%%%%%%%%%%%%%%%%%%%%%%%%%%%%%%%%%%%%%%%%%%%%%%%%%%%%
\section{Detecting critical slowing-down in frequency and time domains\label{sec:Frequency-domain}}
%%%%%%%%%%%%%%%%%%%%%%%%%%%%%%%%%%%%%%%%%%%%%%%%%%%%%%%%%%%%%%%%%%%%%%%%%%%

According to the expression (\ref{eq:CorrFfinal}), as $\epsilon\to0$
and the operating regime of the grid approaches the bifurcation, amplitude
of the fluctuations of the system state vector $x$ grows as $\epsilon^{-3}$.%
\footnote{Note that the actual behavior of the amplitude of fluctuations as
a function of $\epsilon$ depends on the load model, in particular,
on the dependence of the expectation value of fluctuating loads on
frequency $f$ ($\sim{}f^{-2}$ in our case).%
} Simultaneously, the characteristic correlation time of these fluctuations
$\tau_{{\rm corr}}=\frac{a^{T}Db}{\epsilon}$ grows as $\epsilon^{-1}$
at $\epsilon\to{}0$. These two effects explain the phenomenon of
critical slowing-down often observed during large-scale failures of
power grids \cite{Hines2011,Cotilla-Sanchez2012,Podolsky2012,Ghanavati2013}.
The growth of both the amplitude of fluctuations and the correlation
time imply that close to collapse dynamics of the system vector $x$
can be represented as a sequence of relatively long time intervals
with weakly changing values of system variables $x$. However, these
values significantly deviate from the equilibrium ones, $x_{0}$,
determined by the stationarity condition $\dot{x}_{0}=0$. This in
turn (at least partially) explains why it is technically hard to correctly
and rapidly identify the state of the power system close to an unstable
regime using the standard power flow estimated procedures \cite{Wu1977,Klump2000}:
away from the threshold of instability stochastic fluctuations of
the system state vector $x$ are suppressed, while close to this threshold
they are strongly amplified. The same applies to numerical errors
of power flow estimators in operating regimes near bifurcation and
prevents effective convergence of numerical schemes. It is very tempting
to use the very signatures of critical slowing-down as markers of
approach to stability loss \cite{Hines2011,Cotilla-Sanchez2012,Podolsky2012,Ghanavati2013}.
In this Section we shall consider how the critical slowing-down affects
behavior of the autocorrelation function in the frequency domain and
then compare our conclusions to results of numerical simulations. 

The complete dominance of a single mode close to collapse implies
that in the frequency domain expression (\ref{eq:AutoCorrFrequency})
reduces to 
%%%%%%%%%%%%%%%%%%%%%
\begin{equation}
C(f)=\frac{2b(a^{T}{\cal B}a)b^{T}}{(2\pi f)^{2}((a^{T}{\cal D}b)^{2}(2\pi f)^{2}+\epsilon^{2})}+{\cal O}(\epsilon).
\label{eq:AutoCorrFrequencySingleMode}
\end{equation}
%%%%%%%%%%%%%%%%%%%%%
Thus, at very small angular frequencies $2\pi f\ll\frac{\epsilon}{a^{T}{\cal D}b}$
one finds a quadratic behavior of the Fourier-transformed autocorrelation
function (\ref{eq:AutoCorrFrequency}) with $f^{-1}$: $S(f)\approx\frac{2b(a^{T}{\cal B}a)b^{T}}{(2\pi f\epsilon)^{2}}$,
while in the intermediate range of frequencies this behavior is quartic
instead: $S(f)\approx\frac{2b(a^{T}{\cal B}a)b^{T}}{(a^{T}{\cal D}b)^{2}(2\pi f)^{4}}$.
These two types of behavior are rather universal and can be seen even
if the dynamics of the power system does not yet exhibit critical
slowing-down. Closer to the threshold of instability, as $\epsilon\to0$,
the quadratic behavior of (\ref{eq:AutoCorrFrequencySingleMode})
is realized in a smaller and smaller range of frequencies $\tau^{-1}<2\pi f<\frac{\epsilon}{a^{T}{\cal D}b}$.
This in turn leads to a stronger $1/f^{4}$ amplification of the amplitude
of fluctuations at small $f$: the matching point $2\pi f\sim\frac{\epsilon}{a^{T}{\cal D}b}$
between $f^{-4}$ and $f^{-2}$ regimes is reached at smaller and
smaller $f$ , while $\epsilon\to0$. Corresponding amplification
of the amplitude of fluctuations of $x$ particularly noticeable at
small frequencies is a very good marker of approach to stability loss.

In order to check these predictions, we have performed numerical simulations
of the stochastic behavior of IEEE 39 and IEEE 57 test power systems
close to the threshold of instability in both frequency and time domains.
Bifurcation points for both power systems discussed here were first
localized using continuation power flow procedure \cite{Ajjarapu1992,Avalos2009}
implemented in PSAT Toolbox for Matlab \cite{Milano2005} and then
identified more precisely using MATPOWER library for Matlab \cite{Zimmerman2011}.

For the case of the IEEE 39 test power system two chosen values of
the continuation power flow parameter corresponded to the smallest
eigenvalue of the power flow Jacobian $\epsilon\approx0.57$ p.u.
(operating regime relatively far from the threshold of instability)
and $\epsilon\approx0.08$ p.u. (pre-critical operating regime).The
loads were allowed to fluctuate only on the buses 3, 10 and 21 in
order to check the difference in correlations of the state vector
between the nodes with and without fluctuating power loads. On each
of these nodes, a single realization of the active power load fluctuations
was considered (although realizations of $\delta P_{i}$ of course
differed between the nodes). Other parameters of the model were chosen
similarly to \cite{Pai1989,Podolsky2012}, which brings the system
into an overdamped operating regime.

For the case of the IEEE 57 test power system the operational regimes
chosen for simulations corresponded to $\epsilon\approx0.01$ p.u.
(pre-critical regime) and $\epsilon\approx0.17$ p.u. (normal stable
operating regime). Parameters $\alpha_{i}$ were chosen as described
in the Section 2. Inertia constants $H_{i}$ of generators were chosen
according to the relation $H_{i}\approx{}0.04{}P_{i}$, which, as
explained in \cite{Motter2013}, effectively holds for many test power
system models. As usual, the bus $1$ was the slack bus. The active
power loads were allowed to fluctuate only on the nodes $12,28,45$
with characteristic amplitude of fluctuations $\sqrt{{\cal B}}/(2\pi)\approx0.1$
p.u.

\begin{figure}[tbh]
\includegraphics[scale=0.5]{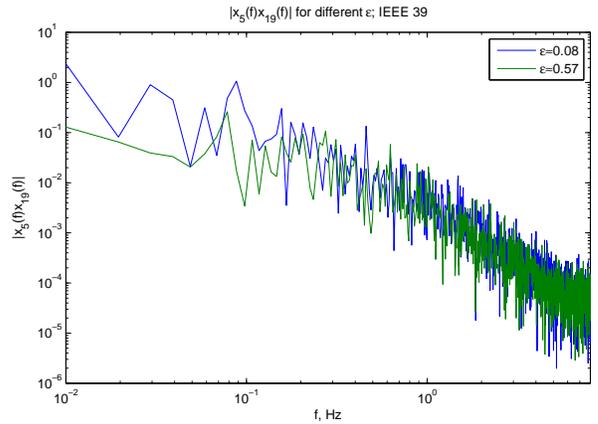} 
\protect\protect\protect\caption{Detecting critical slowing-down in frequency domain; simulations of
IEEE 39. Inter-node correlation function $C(x_{5}(f)x_{19}(f))$,
the load on both nodes $5$ and $19$. The regime $\epsilon\approx0.08$
p.u. is close to the threshold of instability, while the regime with
$\epsilon\approx0.57$ p.u. is relatively far from the threshold.
Amplification of the amplitude of fluctuations at small frequencies
$f$ by more than an order of magnitude is clearly seen. The main
plot uses the log-log scale. The inset represents behavior of the
correlation function at small $f$ using the normal scale. 
\label{fig:IEEE39-freq}}
\end{figure}

\begin{figure}[tbh]
\includegraphics[scale=0.4]{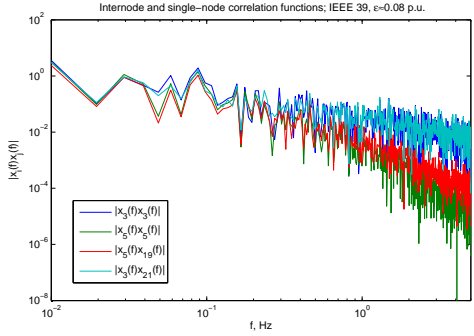} 
\protect\protect\protect\caption{Detecting critical slowing-down in the frequency domain; simulations
of IEEE 39. Comparison of different inter-node and single-node structure
functions in the operating regime close to the threshold of instability.
As can be clearly seen, behavior of different inter-node structure
functions is very similar at small frequencies. The log-log scale
is used for the main plot. The inset represents behavior of the same
correlation functions at small $f$ using the normal scale. The difference
in amplitude of correlation functions at very small $f$ is fully
explained by the difference in magnitudes of the corresponding components
of the vector $b$. 
\label{fig:IEEE39freq2}}
\end{figure}

The results of frequency domain simulations for the IEEE 39 model
are presented on the Figs.\ref{fig:IEEE39-freq}, \ref{fig:IEEE39freq2}.
The $1/f^{4}$ behavior along the relevant interval of frequencies
is clearly seen for both cases of $\epsilon\approx0.57$ p.u. and
$0.08$ p.u., as well as an amplification of the fluctuations at small
frequencies by more than an order of magnitude in the pre-critical
operating regime. The prediction of the theory for the value of the
auto-correlation function at $f\approx0.01$Hz is $\approx{}3$ for
the case $\epsilon\approx0.08$ p.u. and $\approx0.1$ for the case
$\epsilon\approx0.57$ p.u., in a good agreement with the results
of simulation.

As expected, we have found that the low-frequency behavior of different
inter-node/single-node structure functions of the system state vector
$x$ is rather similar irrespective whether the power load is fluctuating
or fixed on the given node. This implies that a single-node autocorrelation
function of the system state vector, in particular, its behavior at
small $f$, can be a good indicator of the approach to the loss of
stability. The behavior at large frequencies differed noticeably,
with inter-node correlations decaying more rapidly with $f$ for nodes
with fixed power loads. The results of simulations of the IEEE 57
test power system are essentially similar. 

\begin{figure}[tbh]
\includegraphics[scale=0.5]{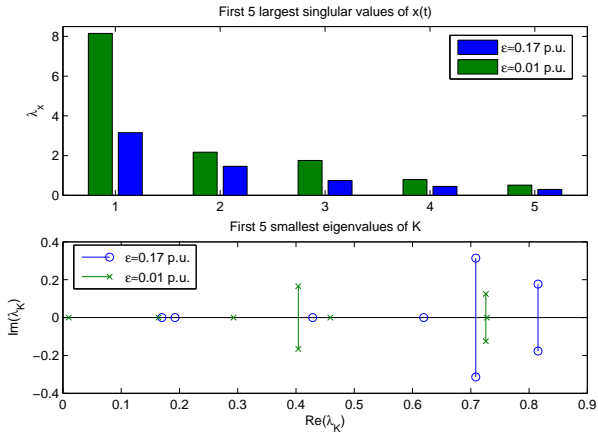} 
\protect\protect\protect\caption{$5$ largest singular values of the system state vector $x$ of the
IEEE 57 power system for pre-critical ($\epsilon\approx0.01$ p.u.)
and sub-critical ($\epsilon\approx0.17$ p.u.) operating regimes.
$8$ smallest eigenvalues of the power flow Jacobian for the same
regimes. 
\label{fig:eigsvd}}
\end{figure}

The results of simulations of the autostructure function $S(|\delta t|)$
in time domain demonstate its amplification at $|\delta t|\sim1-10$
sec. 
%%%%%%%%%%%%%%%%%%%%%%%%%%%%%%%%%%%
%%%%%%%%%%%%%%%%%%%%%%%%%%%%%%%%%
%%%%%%%%%%%%%%%%%%%%%%%%%%%%%%%%%%
The singular value decomposition of the system state vector $x(t)$
shows that there exists a dominating mode for the pre-critical regime
$\epsilon\approx0.01$ p.u., while this is not the case for the stable
operating regime $\epsilon\approx0.17$ p.u., where the contributions
of many different modes to $x$ provide a contribution of the same
order of magnitude into (\ref{eq:CorrFourier}), see Fig. \ref{fig:eigsvd}.
%%%%%%%%%%%%%%%%%%%%%%%%%%%%%%%%%%%
%%%%%%%%%%%%%%%%%%%%%%%%%%%%%%%%%%%
Finally, we would like to emphasize that the contribution of different
nodes into the right eigenvector $b$ of the leading mode relatively
weakly depends on the operating regime. Thus, the dominant nodes of
the grid contributing the most to this mode can be identified well
before the threshold of instability is approached.

%%%%%%%%%%%%%%%%%%%%%%%%%%%%%%%%%%%%%%%%%%%%%%%%%%
\section{Conclusion\label{sec:Conclusion}}
%%%%%%%%%%%%%%%%%%%%%%%%%%%%%%%%%%%%%%%%%%%%%%%%%%

In the present contribution we have studied the phenomenon of critical
slowing-down often observed in power grids close to the threshold
of a large-scale failure such as voltage collapse or a loss of synchrony
\cite{Hines2011,Cotilla-Sanchez2012,Podolsky2012,Ghanavati2013}.
As we have argued, this phenomenon is directly related to (and can
be explained by) a strong amplification of fluctuations of the system
state variables - voltage phases and magnitudes on individual nodes
of the grid - for operating regimes close to collapse. 

We have explained how the phenomenon of critical slowing-down can
be used to effectively detect approach to the threshold of instability.
A technically simple method of detection is based on analysis of single-node
(or inter-node) autostructure functions of the system state vector
in the frequency domain, where approach to the threshold of instability
is characterized by a strong amplification of low frequency part of
the correlation function. It is very important to note that the onset
of critical slowing-down can be detected using limited measurements
of state variables, such as in the situation of incomplete observability
of the smart grid, when effective and precise measurement of inter-area
modes is impossible.

\section*{Acknowledgments}

This work was partially supported by NSF award ECCS-1128437 and MIT/SkTech
seed funding grant.

\end{document}